\documentclass[seceq]{ptptex}

\usepackage{graphicx}

\newcommand{\beq}{\begin{equation}}
\newcommand{\beqa}{\begin{eqnarray}}
\newcommand{\eeq}{\end{equation}}
\newcommand{\eeqa}{\end{eqnarray}}
\newcommand{\siml}{\lesssim}
\newcommand{\simg}{\gtrsim}

\notypesetlogo                       

\title{
A Gamma-Ray Burst/Pulsar for Cosmic Ray Positrons \\
with a Dark Matter-Like Spectrum
}

\author{
Kunihito \textsc{Ioka}
}

\inst{
KEK Theory Center
and the Graduate University for Advanced Studies (Sokendai), 
1-1 Oho, Tsukuba 305-0801, Japan
}

\abst{
We propose that a nearby gamma-ray burst (GRB)
or GRB-like (old, single, and short-lived) pulsar, supernova remnant, or 
microquasar about $10^{5-6}$ years ago
may be responsible for the excesses of cosmic ray
positrons and electrons recently observed in the PAMELA,
ATIC/PPB-BETS, Fermi, and HESS experiments.
We can reproduce the smooth Fermi/HESS spectra
as well as the spiky ATIC/PPB-BETS spectra.
The spectra have a sharp cutoff
that is similar to the dark matter predictions,
sometimes together with a line (not similar),
since high-energy cosmic rays cool fast where
the cutoff/line energy marks the source age.
A GRB-like astrophysical source is expected to have
a small but finite spread in the cutoff/line
as well as anisotropy in the cosmic ray and diffuse gamma-ray flux, 
providing a method for the 
Fermi and future CALET experiments
to discriminate between dark matter and astrophysical origins.
}

\begin{document}

\maketitle

\section{Introduction}
Recent observations by the PAMELA \cite{Adriani:2008zr,Adriani:2008zq} 
and ATIC/PPB-BETS \cite{chang:2008,Torii:2008xu}
experiments have revealed 
the electron and positron excesses 
in the cosmic ray spectrum.
These results indicate the presence of 
nearby sources of electron-positron pairs (less than $1$ kpc away).
Possible candidates include
astrophysical objects such as pulsars
\cite{Kawanaka:2009dk,Shen70,Buesching:2008hr,Hooper:2008kg,
Yuksel:2008rf,chi:1996,zhang:2001,grimani:2007,Profumo:2008ms,Aha:95,Malyshev:2009tw,Grasso:2009ma},
supernova (SN) remnants, \cite{Fujita:2009wk,ShenBerkey68,cowsik79,boulares89,erlykin02,pohl98,Kobayashi:2003kp,
Shaviv:2009bu,Hu:2009bc,Blasi:2009hv,Blasi:2009bd,Mertsch:2009ph,Biermann:2009qi}
or microquasars \cite{Heinz:2002qj},
or dark matter annihilations/decays. \cite{Asano:2006nr,ArkaniHamed:2008qn,
Bergstrom:2008gr,Cirelli:2008jk,Chen:2008yi,Chen:2008qs,Chen:2008dh,Cholis:2008wq,Cirelli:2008pk,Hamaguchi:2008ta,Hisano:2008ah,Hisano:2004ds,Ishiwata:2008cv,Hall:2008qu,Zhang:2008tb,MarchRussell:2008tu,Hooper:2008kv}
Instead, we might be observing
the propagation effects
\cite{Delahaye:2007fr,Cowsik:2009ga,Katz:2009yd,Stawarz:2009ig,Schlickeiser:2009qq}
or the proton contamination
\cite{Fazely:2009jb,Schubnell:2009gk,Israel:2009}.

The ATIC/PPB-BETS excess has a possible cutoff
at $\varepsilon_e \sim 600$ GeV,
which might fix the dark matter mass.
From the astrophysical viewpoint, the cutoff implies
a single or at least a few sources since many sources
usually broaden the cutoff.
The source age should be less than
$10^{6{\rm -}7}$ years
because electrons lose energy through synchrotron
and inverse Compton processes,
suggesting the Galactic rate of
$\sim (10\ {\rm kpc}/1\ {\rm kpc})^2/10^{6{\rm -}7}\ {\rm yr} 
\sim 1/10^{4{\rm -}5}$ yr,
i.e., $\sim 10^2$--$10^3$ times rarer than SNe.
This ratio $\sim 10^2$--$10^3$ is comparable to that of the energy density 
between cosmic ray nuclei and positrons.
Therefore, the electron-positron source may also produce
a huge energy $\sim 10^{50}$ erg like an SN
that releases $\sim 10^{50}$ erg for providing cosmic ray nuclei.

In this paper, we propose a new possibility
that a nearby ($d \sim 1$ kpc) gamma-ray burst (GRB) 
or GRB-like pulsar/SN remnant/microquasar
about $t_{\rm age}\sim 10^{5-6}$ years ago
may be responsible for the PAMELA and ATIC/PPB-BETS excesses,
and predict a sharp spectral cutoff 
that is similar to the dark matter predictions,
in addition to a possible line.
GRBs are the most luminous objects in the universe \cite{Meszaros:2006rc,Zhang:2007nka},
brief ($\sim 1$--$100$ sec) bursts of 
high-energy ($\sim 0.1$--$1$ MeV) photons
appearing at random in the sky
about 1000 times per year 
(corresponding to the collimation-corrected local rate 
$\sim 1/10^{5{\rm -}6}$yr/galaxy
\cite{Guetta:2006gq}).
Thanks to the discovery of afterglows (long-lasting counterparts
in longer wavelengths) and SNe,
it is widely accepted that (long) GRBs are
associated with the deaths of massive stars. 
The central core of a massive star gravitationally collapses into a black hole or 
neutron star, which somehow launches a collimated outflow (jet)
and produces GRBs and afterglows with a typical true energy 
of $\sim 10^{51}$ erg.
GRBs could emit a significant amount of energy that is comparable 
to the main MeV photons
into GeV-TeV gamma-rays, as observed by the Fermi satellite
\cite{Abdo09},
and eV photons (so-called optical flashes)
as in the famous naked-eye GRB 080319B \cite{Racusin:2008pd}.

Very recently, 
the Fermi Large Area Telescope has measured the electron spectrum
up to $\sim 1$ TeV, which is very smooth $\sim \varepsilon_e^{-3}$
without any spectral peak as reported by ATIC/PPB-BETS.
The HESS collaboration also provides the electron spectrum
\cite{Collaboration:2008aa,Aharonian:2009ah},
which is consistent with the Fermi result
and appears to show a steepening above $\sim 1$ TeV.
The differences between ATIC/PPB-BETS and Fermi/HESS are still 
controversial because Fermi removes a significant fraction of electrons
above $\sim 300$ GeV
to avoid the $\sim 10^{3{\rm -}4}$ times larger hadron contamination 
and reconstructs the real flux by the Monte Carlo simulations
\cite{Moiseev:2007js,Israel:2009},
while the statistical errors of ATIC are much worse than those of Fermi.
Therefore, we discuss each case separately and show that
a GRB/pulsar model with sightly different parameters
may reproduce the Fermi/HESS smooth spectra as well.

In this paper, we try not to specify a source class
to make the discussion as model-independently as possible.
We consider a GRB-like astrophysical source,
denoted as GRB/Pulsar for short,
which produces electron-positron pairs from a compact region
in a short timescale compared with its age.
If the source has these properties,
we can apply the results to pulsars, SN remnants, microquasars, GRBs, etc.
The exception is \S\ref{sec:GRB}, which is devoted to GRB models.

\section{ATIC/PAMELA excess from an astrophysical source}
\label{sec:ATIC}
Let us first consider the most simple model
that a GRB/pulsar produces 
electron-positron pairs with energy $E_{e^+}\simeq E_{e^-}$
at a distance $d$ from Earth at time $t_{\rm age}$ ago,
assuming that the pairs have a power-law spectrum.
The observed spectrum after propagation
is obtained by solving the diffusion equation,
\beqa
\frac{\partial}{\partial t} f
=K(\varepsilon_e) \nabla^2 f
+\frac{\partial}{\partial \varepsilon_e} [B(\varepsilon_e) f]
+Q,
\label{eq:diff}
\eeqa
where $f(t,\vec{x},\varepsilon_e)$
is the distribution function of particles at time $t$
and position $\vec{x}$ with energy $\varepsilon_e$.
The flux at $\vec{x}$ is given by 
$\Phi(t,\vec{x},\varepsilon_e)=(c/4\pi) f(t,\vec{x},\varepsilon_e)$
[m$^{-2}$ s$^{-1}$ sr$^{-1}$ GeV$^{-1}$].
We adopt the diffusion coefficient 
$K(\varepsilon_e)=K_0 (1+\varepsilon_e/3\ {\rm GeV})^\delta$
with $K_0=5.8 \times 10^{28}$ cm$^2$ s$^{-1}$ 
and 
$\delta=0.6$
that is consistent with the boron/carbon ratio
according to the latest GALPROP code,
and the energy loss rate $B(\varepsilon_e)=b \varepsilon_e^2$
with $b=10^{-16}$ GeV$^{-1}$ s$^{-1}$ 
via synchrotron and inverse Compton
\cite{Strong:2004de,Baltz:1998xv,Moskalenko:1997gh}.

In the limit of a single burst from a point source
with a power-law spectrum
$Q(t,\vec{x},\varepsilon_e)=Q_0 \varepsilon_e^{-\alpha} 
\delta(\vec{x}) \delta(t)$ 
up to $\varepsilon_e<\varepsilon_{\max}$,
the diffusion Eq.~(\ref{eq:diff}) has an analytical solution as
\cite{Atoyan:1995}
\beqa
f=\frac{Q_0 \varepsilon_e^{-\alpha}}{\pi^{3/2} d_{\rm diff}^3}
(1-b t \varepsilon_e)^{\alpha-2}
e^{-(d/d_{\rm diff})^2},
\label{eq:fsol}
\eeqa
where $\varepsilon_e<(b t +1/\varepsilon_{\max})^{-1}
<\varepsilon_{\rm cut}=(b t)^{-1}$ (otherwise $f=0$)
and
\beqa
d_{\rm diff} \simeq 2 \sqrt{K(\varepsilon_e) t 
\frac{1-(1-\varepsilon_{e}/\varepsilon_{\rm cut})^{1-\delta}}
{(1-\delta)\varepsilon_{e}/\varepsilon_{\rm cut}}
}.
\label{eq:ddiff}
\eeqa
The physical picture is that cosmic rays below 
$\varepsilon_e \siml \varepsilon_{\rm cut}$
diffuse out almost uniformly within a radius
$d_{\rm diff} \sim  2 \sqrt{K(\varepsilon_e) t}$.

\begin{figure}
\centerline{\includegraphics{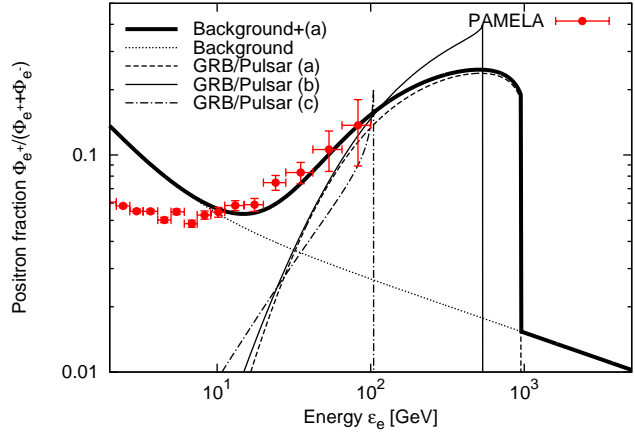}}
\caption{
Positron fraction $\Phi_{e^+}/(\Phi_{e^-}+\Phi_{e^-})$
resulting from a GRB/pulsar [(a), (b), (c)]
and secondary positrons produced by
the collisions of cosmic ray nuclei with interstellar medium (ISM)
[Background],
compared with the PAMELA data.
The fit is well
and the spectrum has a cutoff at
$\varepsilon_e  \sim \varepsilon_{\rm cut}$
in Eq.~(\ref{eq:emax}).
We adopt $(t_{\rm age},E_{e^+},\alpha)=
(3 \times 10^{5}\ {\rm yr}, 0.7 \times 10^{50}\ {\rm erg}, 2.2),
(5.6 \times 10^{5}\ {\rm yr}, 1.2 \times 10^{50}\ {\rm erg}, 1.6)$
and $(3 \times 10^{6}\ {\rm yr}, 5 \times 10^{50}\ {\rm erg}, 1.5)$
for (a), (b), and (c), respectively,
where a GRB/pulsar at $d=1$ kpc from Earth
at time $t_{\rm age}$ ago
produces electron-positron pairs with energy 
$E_{e^+}=E_{e^-}$
and spectral index $\alpha$ up to $\varepsilon_{\max}=10$ TeV.
A GRB/pulsar (a) fits the Fermi/HESS data,
while a GRB/pulsar (b) fits the ATIC/PPB-BETS data well in Fig.~\ref{fig:flux}.
A GRB/pulsar (c) is an older one.
Note that 
the solar modulation is important below $\sim 10$ GeV.
}
\label{fig:e+frac}
\end{figure}

\begin{figure}
\centerline{\includegraphics{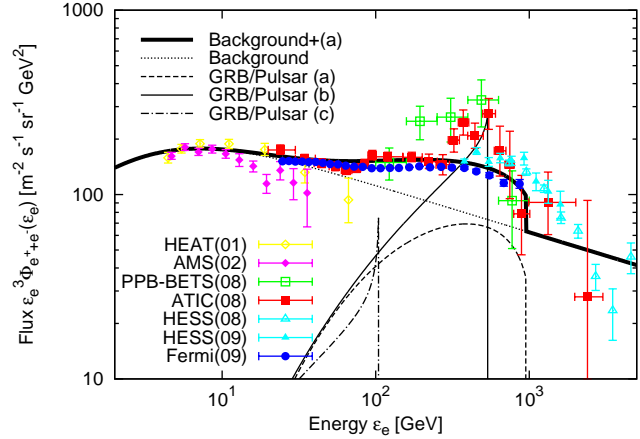}}
\caption{
Electron plus positron flux
from a GRB/pulsar [(a), (b), (c)]
and the primary plus secondary background,
compared with the data.
A GRB/pulsar (a) fits the Fermi/HESS data,
while a GRB/pulsar (b) fits the ATIC/PPB-BETS data well.
A GRB/pulsar (c) is an older one.
We adopt the same parameters as in Fig.~\ref{fig:e+frac}.
The spectrum has a cutoff at
$\varepsilon_e =\varepsilon_{\rm cut}$
in Eq.~(\ref{eq:emax}).
The primary background is conventionally attributed to SN remnants.
Note that 
the solar modulation is important below $\sim 10$ GeV.
}
\label{fig:flux}
\end{figure}

In Figs.~\ref{fig:e+frac} and \ref{fig:flux},
we show the positron fraction 
and the electron plus positron flux
resulting from a GRB/pulsar and background. \footnote{
For the background in all the figures,
we adopt the fitting functions
in Ref. \citen{Baltz:1998xv}
by reducing the primary electron flux by $30\%$
because the fitting functions provide a larger flux than the Fermi and ATIC data
below $\sim 100$ GeV
even without other contributions,
while the secondary flux is fixed by 
the observed cosmic ray nuclei \cite{Moskalenko:1997gh}.
We note that the fitting functions are not reliable 
far above $\sim 1000$ GeV for primary electrons and
$\sim 100$ GeV for the secondary electrons and positrons.
}
We can see that
the PAMELA and ATIC/PPB-BETS excesses
can be reproduced well if
a GRB/pulsar produces electron-positron pairs with
energy $\sim 10^{50}$ erg and a power-law spectral index 
$\alpha \sim 1.6$--$2.2$
at $d \sim 1$ kpc from Earth at time $t_{\rm age} \sim 6 \times 10^5$ yr ago.
The chance probability of having such a GRB
is $t_{\rm age}/10^{5{\rm -}6}\ {\rm yr}/(10\ {\rm kpc}/1\ {\rm kpc})^2 
\sim 0.6$--$6\%$,
which is not too bad.
Otherwise, a pulsar/SN remnant/microquasar per $6$--$60$ SNe may be responsible.

Interestingly, the electron and positron spectra
in Figs.~\ref{fig:e+frac} and \ref{fig:flux}
have a sharp cutoff that is very similar to the dark matter predictions
\cite{Hall:2008qu,Chen:2008fx}
in addition to a line at energy,
\beqa
\varepsilon_{\rm cut}=\frac{1}{bt} \simeq
300 \left(\frac{10^6\ {\rm yr}}{t_{\rm age}}\right) {\rm GeV}.
\label{eq:emax}
\eeqa
This is because the energy loss time
via synchrotron and inverse Compton 
is short for high-energy cosmic rays by $\varepsilon_e$.
Then, after time $t_{\rm age}$,
all the electrons above $\varepsilon_{\rm cut}$ cool to 
the cutoff energy $\varepsilon_{\rm cut}$
with no electrons above $\varepsilon_{\rm cut}$.
Independently of the maximum energy, all the electrons above
$\varepsilon_{\rm cut}$ lose their energies during propagation.
A line or cusp is produced if the source spectrum has 
$\alpha<2$, although the numbers of electrons and positrons
remain finite and constant \cite{Longair:1994}.
Note that the electron and positron lines 
produced by the dark matter are smeared out 
because the observed electrons and positrons are created 
at different times having different line energies due to cooling.
Note also that 
only the direct annihilation or two-body decay into electrons and positrons
can produce a sharp cutoff in dark matter models.

If the ATIC/PPB-BETS cutoff is due to the cooling,
we can estimate the necessary energy 
only from observations as
$E_e \sim \varepsilon_e^2 (4\pi/c) \Phi_e(\varepsilon_e)
4\pi d_{\rm diff}^3/3
\sim 8 \times 10^{48} {\rm erg}$
$(\varepsilon_e/600\ {\rm GeV})^{-1.2+3\delta/2}
(K_0/2 \times 10^{28}\ {\rm cm}^2\ {\rm s}^{-1})^{3/2}$.
This can be yielded by a GRB, SN remnant,
a 10 msec pulsar with a rotational energy $\sim 10^{50}$ erg,
or a microquasar (a black hole with a disk and jet) that has
the Eddington luminosity $\sim 10^{38}$ erg s$^{-1}$ for $\sim 10^5$ yr.

\section{Fermi/PAMELA excess from an astrophysical source}
In this section, we move on to the Fermi/HESS electron spectra,
which have no spectral peak as reported
by ATIC/PPB-BETS (see \S\ref{sec:ATIC} and Fig.~\ref{fig:flux}).
Since the differences between ATIC/PPB-BETS and Fermi/HESS are
still much debated, we discuss each case separately.

In Figs.~\ref{fig:e+frac} and \ref{fig:flux}, 
we also show a GRB/pulsar model (a) that can reproduce
the Fermi/HESS smooth data as well as the PAMELA data
without producing the ATIC/PPB-BETS peak.
Interestingly, the model parameters are relatively similar
to those for the ATIC/PPB-BETS data,
i.e.,
the age is just slightly lower
($6\times 10^5$ yr $\to$ $3\times 10^5$ yr)
and the spectral index is softer
($1.6 \to 2.2$),
where $\alpha=2$ is the boundary
between the smooth and spiky spectra in Eq.~(\ref{eq:fsol}).
We still have a cutoff at $\varepsilon_e =\varepsilon_{\rm cut}$
in Eq.~(\ref{eq:emax}),
while we have no line for $\alpha > 2$
with Eqs.~(\ref{eq:fsol}) and (\ref{eq:emax}).
The cutoff may be relevant to the steepening
observed by HESS around $\sim 1$ TeV,
although we need more data to claim the presence of the cutoff.
(If the steepening continues to higher energy, 
the background would also have a cutoff.)

Considering an older source, for which the chance probability 
becomes higher,
we may fit the PAMELA data, leaving the electron data to other sources
[see case (c) in Figs.~\ref{fig:e+frac} and \ref{fig:flux}].
However, if the electron spectrum is as smooth as the Fermi data, 
it may be difficult to hide the peak under the other contributions
(Note that a condition $\alpha < 2$ is necessary 
to fit the PAMELA data using a single 
old source.)
Thus, a single GRB-like source only for the PAMELA data is unlikely.
If the PAMELA excess is caused by a single source,
the source should also produce the Fermi excess 
[see case (a) in Figs.~\ref{fig:e+frac} and \ref{fig:flux}]
or the source should have a long duration $\Delta t$ 
to have a round peak
(see \S\ref{sec:dis}).
Otherwise, multiple sources may be involved 
to make the spectrum smooth \cite{Kawanaka:2009dk}.

\section{Possible models}\label{sec:GRB}
The PAMELA, ATIC/PPB-BETS, and Fermi/HESS excesses
can be reproduced well if
a GRB/pulsar produces electron-positron pairs with
energy $\sim 10^{50}$ erg and a power-law spectral index 
$\alpha \sim 1.6$--$2.2$.
The next question is how to produce such pairs.
Because we have several discussions for a pulsar/SN remnant/microquasar
\cite{chi:1996,Heinz:2002qj,Fujita:2009wk},
we concentrate here on the GRB case.

The observed GRB spectrum
has a power-law shape with
a significant fraction of energy 
above the pair production threshold.
Thus, it is plausible that the energy 
$\sim 10^{50}$ erg ($\sim 3$--$10\%$ of the total energy)
goes into pairs via $\gamma \gamma \to e^+ e^-$.
Pairs may be created
in the outflowing jet that produces gamma-rays
\cite{Ioka:2007qk,Rees:2004gt}.
As the jet expands, the pair annihilations freeze out,
so that the pair-loaded jet could yield
the right amount of pairs
if pairs can escape into the ISM as cosmic rays.
Note that 
the pair budget may be diagnosed
by observing the blue-shifted pair-annihilation line and cutoff 
with the Fermi satellite \cite{Murase:2007ya}.

Alternatively, pairs could be created far outside a radius 
$R_e > R_c \sim {3 E_e}/{4\pi n m_p K^2}$
$\sim 10^{17}\ {\rm cm}$
$\left({E_e}/{10^{50}\ {\rm erg}}\right)$
$\left({n}/{1\ {\rm cm}^{-3}}\right)^{-1}$
$(K/10^{28}\ {\rm cm}^2\ {\rm s}^{-1})^{-2}$
where an adiabatic cooling \cite{zhang:2001} is not effective 
since the diffusion time
$t_{\rm diff} \sim R_e^2/K$ becomes shorter than
the expansion time
$t_{\rm exp} \sim R_e/v_e 
\sim R_e/(3 E_e/4 \pi R_e^3 n m_p)^{1/2}$
of the ISM with number density $n$.
For efficient pair creation,
the target photon density should be high,
$n_t \simg (\sigma_T R_e)^{-1} \sim 10^{6}\ {\rm cm}^{-3}
(R_e/10^{18}\ {\rm cm})^{-1}$,
which corresponds to the (isotropic) total energy,
\beqa
E_t \sim \frac{4\pi}{3} R_e^3 \varepsilon_t n_t 
\simg 10^{49}\ {\rm erg}
\left(\frac{\varepsilon_t}{1\ {\rm eV}}\right)
\left(\frac{R_e}{10^{18}\ {\rm cm}}\right)^2.
\label{eq:targetph}
\eeqa
A GRB itself may provide target photons,
for example, via the dust scattering
of $\sim 1$ eV photons from an optical flash
associated with a GRB \cite{Waxman:1999rm,Esin:2000kz}.
Most optical photons can become target photons 
because the scattering optical depth may be high
such as in a molecular cloud
(the birth place of the GRB progenitors).
An optical flash can also start during the GRB.
The dust out to a distance of $10^{18}$--$10^{20}$ cm
may be destroyed by sublimation 
\cite{Waxman:1999rm}, producing target photons selectively outside.
Photons absorbed by dust are reradiated in the infrared 
$\varepsilon_t \sim 0.1$ eV, also yielding target photons.
The dust scattering/absorption may explain
the puzzling paucity of optical flashes compared with the theoretical predictions
\cite{Meszaros:2006rc}.

Then, we have very ample target photons with energy 
$\varepsilon_t \sim 0.1$--$1$eV
(i.e., Eq.~(\ref{eq:targetph}) is satisfied and 
the optical depth to the pair creation is so high)
that almost all the GRB photons with energy 
\beqa
\varepsilon_{\gamma} > \varepsilon_{\rm peak} \sim
(m_e c^2)^2/\varepsilon_t \sim 1\ {\rm TeV}
\eeqa
can turn into electron-positron pairs. \footnote{
The Klein-Nishina effect suppresses
the pair creation by very energetic photons.}
Although it is currently unknown
whether GRBs emit TeV photons,
the previous data are consistent with a simple power-law extending
up to the TeV region without any cutoff \cite{Albert:2006jx}.
For a typical power-law index $\sim 2.2$,
the created TeV pairs have
$\sim (200\ {\rm keV}/1\ {\rm TeV})^{0.2} \sim 5\%$ of the total GRB energy,
typically $\sim 10^{50}$ erg (collimation-corrected).
Note that the high-energy photons
do not interact with target photons before scattering
because they are beamed into the same direction in the lab frame
(i.e., the high Lorentz factor of GRBs allows high-energy photons to escape).

After being randomized by magnetic fields,
the TeV pairs are exposed to the afterglow,
upscattering $\varepsilon_a \sim 10^{-3}$--$1$ eV afterglow photons 
into GeV-TeV photons via inverse Compton,
which create GeV-TeV pairs with other afterglow photons
through $\gamma \gamma \to e^+ e^-$,
where the optical depth is $\tau \sim \sigma_T c t U_a/\varepsilon_a$
$\sim 1\ (E_a/10^{49}{\rm erg})$ 
$(R_e/10^{18}{\rm cm})^{-2}$
$(\varepsilon_a/1{\rm eV})^{-1}$.
The total energy of GeV-TeV pairs is mainly provided by
pairs initially above the cooling cutoff
$\varepsilon_{\rm cut}^{a} = 1/b_{a} t$,
where $b_{a}=4 \sigma_T c U_a/3 m_e^2 c^4$
is the energy loss rate via inverse Compton
of afterglow photons with the energy density $U_a$.
Since the afterglow luminosity typically decays as $\sim t^{-1}$,
i.e., $U_a \sim E_a/4\pi R_e^2 c t$,
the cutoff energy is almost time-independent at
\beqa
\varepsilon_{\rm cut}^{a} \sim \frac{3 \pi R_e^2 m_e^2 c^4}{\sigma_T E_a}
\sim 600\ {\rm GeV} \left(\frac{R_e}{10^{18}\ {\rm cm}}\right)^2
\left(\frac{E_a}{10^{49}\ {\rm erg}}\right)^{-1},
\label{eq:emaxa}
\eeqa
where $E_a$ is the total energy of afterglow photons.
Since the cutoff energy $\varepsilon_{\rm cut}^{a}$
is not so different from the initial peak energy 
$\varepsilon_{\rm peak} \sim 1$TeV,
the created GeV-TeV pairs 
have comparable energy $\sim 10^{50}$ erg with initial TeV pairs.
The photons upscattered by an electron or positron basically copy
the afterglow spectral shape,
which has a range of power-law index $1$--$2.5$
depending on the time and spectral segments \cite{Meszaros:2006rc}.
Thus, a spectral index $\alpha \sim 1.8 \pm 0.7$ is also possible
although detailed calculations are needed to obtain the final spectrum.
Note that in some parameter space,
$\varepsilon_{\rm cut}^{a}<\varepsilon_{\rm cut}$ 
in Eqs.~(\ref{eq:emax}) and (\ref{eq:emaxa}),
so that the cutoff may be shaped by the afterglow, not the source age.

\section{Single or multiple? Leptonic or hadronic?}

\begin{figure}[t]
\centerline{\includegraphics{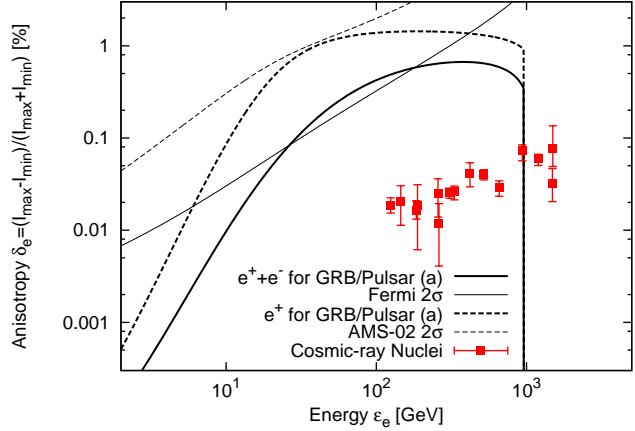}}
\caption{
Dipole anisotropy of the
electron plus positron flux (thick solid line)
and the positron flux (thick dashed line)
for the GRB/pulsar model (a) in Figs.~\ref{fig:e+frac} and
\ref{fig:flux},
compared with sensitivities of 
the Fermi satellite for the electron plus positron flux 
(thin solid line),
the future AMS-02 experiment 
for the positron flux (thin dashed line),
and the observed anisotropy of cosmic ray nuclei
$\delta_N \sim 0.06\%$ \cite{Guillian:2005wp} (filled square).
}
\label{fig:aniso}
\end{figure}

It is important to determine whether the source is single
or multiple.
To answer this question, an anisotropy measurement could 
be useful. \cite{Buesching:2008hr,MaoShen72}
In Fig.~\ref{fig:aniso}, we show the expected anisotropy 
of electron and positron fluxes
\beqa
\delta_e= \frac{I_{\rm cut}-I_{\min}}
{I_{\rm cut}+I_{\min}}
=\frac{3 K |\nabla f|}{c f},
\eeqa 
for the GRB/pulsar model (a) in Figs.~\ref{fig:e+frac} and \ref{fig:flux}.
The anisotropy is larger than that of the observed cosmic ray nuclei
$\delta_N \sim 0.06\%$ \cite{Guillian:2005wp},
so that the anisotropy is in principle detectable,
not to be disturbed by the local magnetic structure.
The Fermi and upcoming AMS-02 experiments
may be able to detect the anisotropy, while
the actual measurement should be challenging 
and also model-dependent,
e.g., the GRB/pulsar model (b) in Figs.~\ref{fig:e+frac} and \ref{fig:flux} 
predicts the anisotropy below the sensitivities
(not shown in Fig.~\ref{fig:aniso})
mainly because its older age leads to a smoother distribution via diffusion.
Once an anisotropy is detected, it would support
a single source model, not a multiple source model.
In Fig.~\ref{fig:aniso}, we use 
a condition $\delta_e \simg 2 \sqrt{2/N_e}$ for 
the $2 \sigma$ detection where
the numbers of electrons and positrons
are obtained from
the thick solid lines in Figs.~\ref{fig:e+frac} and \ref{fig:flux}.
The Fermi satellite will detect
$N_{e^++e^-} \sim 10^8$ electrons and positrons
above 10 GeV in 5 years without charge separation, \cite{Moiseev:2007js}
whereas the AMS-02 experiment will detect 
$N_{e^+} \sim 10^6$ positrons above 10 GeV in 3 years.

We should be careful about the anisotropy arguments 
since the strong isotropy of the cosmic ray nuclei 
is not fully understood yet.
In the diffusion picture, the anisotropy is
connected with the diffusion coefficient
as $\delta_N \sim 3K/c H \sim 0.1 \%\ 
(K/6 \times 10^{28}\ {\rm cm}^2\ {\rm s}^{-1})
(H/2\ {\rm kpc})^{-1}$,
where $H$ is the characteristic height of cosmic rays,
which is consistent with the
residence time of cosmic rays in the Galactic disc
$\tau \sim H^2/2K \sim 1 \times 10^7\ {\rm yr}\ 
(K/6 \times 10^{28}\ {\rm cm}^2\ {\rm s}^{-1})^{-1}
(H/2\ {\rm kpc})^2$ \cite{wentzel74,cesarsky80}.
However, this anisotropy in the diffusion picture 
is somewhat larger than the observed one,
and is also energy-dependent in contrast with
the observed weak energy dependence between
$\sim 1$ and $\sim 10^{2}$ TeV \cite{Guillian:2005wp}.
One possibility is that the anisotropy is suppressed
by nearby sources that contribute in the opposite direction
\cite{Erlykin:2006ri,Ptuskin:2006}.
This may be conceivable because
there is no finite second moment in the distribution
so that the decisive causes of the fluctuations are the nearby sources
\cite{lee79,lagutin95}.
Second, the anisotropy could be induced by the local interstellar
magnetic field, \cite{Amenomori:2007,Battaner:2009zf}
although this model cannot explain
the weak energy dependence of the anisotropy
in a straightforward way.
The heliospheric magnetic field may also be
responsible for the decreasing anisotropy below $\sim 1$ TeV
\cite{Munakata:2009di}.
Finally, only the Compton-Getting effect
cannot account for the anisotropy
\cite{Amenomori:2006bx}.
In any case, the same reason for the suppression of
the anisotropy of the cosmic ray nuclei may also affect 
the electron and positron anisotropies.

If a single event has a significant contribution to 
the positron and electron fluxes, multiple events
of the same type may also be important for the flux.
Given the event rate of the sources $R_{s}$ [s$^{-1}$ cm$^{-2}$],
we may estimate the total average flux from multiple events as
\beqa
f_{\rm ave} \simeq \int_{0}^{1/b \varepsilon_e} dt
\int_{0}^{d_{\rm diff}} dr \, 2\pi r f R_{s},
\eeqa
where $f=f(t,d=r,\varepsilon_e)$ is the single event contribution 
in Eq.~(\ref{eq:fsol}) and non-negligible for $t < 1/b\varepsilon_e$
(cooling time) and $d=r < d_{\rm diff}$ (diffusion radius),
and we assume that the sources are born in a thin disk.
Approximating $f \propto \varepsilon_e^{-\alpha}/d_{\rm diff}^3$ for 
$t < 1/b\varepsilon_e$ and $d=r < d_{\rm diff}$
and $d_{\rm diff} \sim 2 \sqrt{K(\varepsilon_e) t} 
\propto \varepsilon_e^{\delta/2} t^{1/2}$,
we have
\beqa
f_{\rm ave} \propto \varepsilon_e^{-\alpha -\delta/2 -1/2},
\eeqa
which is $\sim \varepsilon_e^{-3}$ for $(\alpha,\delta)=(2.2,0.6)$
and $\sim \varepsilon_e^{-2.4}$ for $(\alpha,\delta)=(1.6,0.6)$.
Therefore, the multiple contributions tend to make the spectrum softer.
We may fit the observed spectrum by using a harder spectral index $\alpha$
accordingly [see Ref. \citen{Kawanaka:2009dk} for details].
In this case, the high-energy cutoff is produced by the latest event.

Whether the antimatter origin is hadronic or leptonic
is also an important problem
(e.g., the pulsar model is leptonic and the SN remnant model 
with $pp$ interactions is hadronic.)
Fujita et al. \cite{Fujita:2009wk} first pointed out that 
the hadronic models predict
an antiproton excess above $\sim 100$ GeV,
which will be probed with PAMELA and future AMS-02
[see also Ref. \citen{Blasi:2009bd}].
The secondary nuclei such as the boron-to-carbon and titanium-to-iron ratio
would also be an interesting probe \cite{Mertsch:2009ph}.


\section{Discussion and summary}\label{sec:dis}
We have proposed that a nearby gamma-ray burst (GRB)
or GRB-like (old, single, and short-lived) pulsar/SN remnant/microquasar
about $10^{5-6}$ years ago may be responsible for the excesses of cosmic ray
positrons and electrons recently observed in the PAMELA,
ATIC/PPB-BETS, and Fermi/HESS experiments.
Such a scenario appears extreme but still consistent with
the current observations.
In particular, a GRB/pulsar model can reproduce 
the smooth Fermi/HESS spectra as well as the spiky ATIC/PPB-BETS spectra
by slightly changing the parameters (see Fig.~\ref{fig:flux}).

Although such a burstlike scenario was discussed previously,
\cite{Atoyan:1995,ShenBerkey68,Shen70}
it is the first to argue the similarities (e.g., sharp cutoff) 
and differences (e.g., cutoff width) between the 
astrophysical and dark matter scenarios.\footnote{
Our work was carried out independently of 
Profumo, \cite{Profumo:2008ms} who also pointed out 
that an astrophysical source (pulsar) can produce a sharp spectral cutoff.}
In particular, we propose a new method to discriminate models
by using the cutoff width (see below).
This new viewpoint arises from the confrontation 
between astrophysical and dark matter models as well as the 
developments of experiments with fine spectral resolutions, 
both of which did not exist previously.
The GRB model is also a new one, and the markedly improved quality of 
current data would allow us to reconsider the problem of whether a 
GRB-like source can explain the data or not.

The spectral cutoff and line in Figs.~\ref{fig:e+frac} and \ref{fig:flux}
should have a finite dispersion under realistic circumstances,
in contrast with the dark matter origin \cite{Chen:2008fx}.
We may be able to discriminate models by observing the cutoff shape (or width)
since 
the future CALET experiment has a resolution 
better than a few $\%$ ($>100$ GeV) \cite{torii:2006,torii:2008}.

Since Eq.~(\ref{eq:emax}) yields
${\Delta \varepsilon_{\rm cut}}/{\varepsilon_{\rm cut}}
=-{\Delta b}/{b}-{\Delta t}/{t}$,
the dispersion arises from 
(a) the fluctuation of the energy loss rate $\Delta b$
due to the difference of starlight and magnetic fields by location
and (b) the duration of the source $\Delta t$.
To estimate $\Delta b$, we assume
that the energy loss rate fluctuates by $\delta b$ over the scale $d_{b}$.
Cosmic rays travel a distance $c t_{\rm age}$
and pass through $N_b \sim c t_{\rm age}/d_b$ patches,
averaging the fluctuations as
$\Delta b \sim \delta b/\sqrt{N_b}$.
Then, we have
\beqa
\left(\frac{\Delta \varepsilon_{\rm cut}}{\varepsilon_{\rm cut}}\right)_{\Delta b}
\sim 6\% \
\left(\frac{\delta b}{b}\right)
\left(\frac{d_b}{1\ {\rm kpc}}\right)
\left(\frac{t_{\rm age}}{10^6\ {\rm yr}}\right)^{-1/2},
\label{eq:dcutdb}
\eeqa
which may be detectable if 
the starlight and magnetic fields differ by $\delta b/b \sim 1$
over the disk thickness $d_b \sim 1$ kpc
[see also Ref. \citen{Malyshev:2009tw}].
As for the duration effect,
GRBs are too short, but
a pulsar with magnetic field $B$ and initial rotation period $P_0$
has a spin-down duration
$\Delta t \sim 3c^3 I/B^2 R_{*}^6 \Omega_0^2
\sim 6 \times 10^3\ {\rm yr}$ 
$(B/10^{12}\ {\rm G})^{-2} (P_0/10\ {\rm ms})^2$,
whereas a microquasar has an active time
$\Delta t \sim 10^{50}\ {\rm erg}/L \sim 10^{5}\ {\rm yr}$
$(L/10^{38}\ {\rm erg}\ {\rm s}^{-1})^{-1}$,
yielding 
\beqa
\left(\frac{\Delta \varepsilon_{\rm cut}}
{\varepsilon_{\rm cut}}\right)_{\Delta t}
 \sim 
 10\% \left(\frac{\Delta t}{10^{5}\ {\rm yr}}\right) 
\left(\frac{t_{\rm age}}{10^{6}\ {\rm yr}}\right)^{-1}.
\eeqa
[See Ref. \citen{Kawanaka:2009dk} for a more detailed discussion.]

The line shape in Figs.~\ref{fig:e+frac} and \ref{fig:flux} is 
$\propto (1-\varepsilon_e/\varepsilon_{\rm cut})^{\alpha-2}$
from Eqs.~(\ref{eq:fsol}) and (\ref{eq:emax}).
For an energy resolution $\Delta \varepsilon_e/\varepsilon_e=\lambda$,
the flux is enhanced at the line by a factor of
$\lambda^{\alpha-2}/(\alpha-1)$ 
($\sim 2.7$ for $\lambda=2\%$ and $\alpha=1.8$).
As $\varepsilon_e<(b t +1/\varepsilon_{\max})^{-1}
=(1/\varepsilon_{\rm cut}+1/\varepsilon_{\max})^{-1}$,
there is no divergence in the line.

Note that there are still some uncertainties in the diffusion coefficients.
The change in the diffusion coefficients can be
adjusted by slightly changing the model parameters.
The smaller $K$ makes the diffusion length $d_{\rm diff}$ smaller, and
the particle density inside that radius becomes higher, being proportional 
to $d_{\rm diff}^{-3}$. For different $\tilde{K}$
instead of $K$, we can apply our results by rescaling the distance 
and energy as 
$d\rightarrow d\sqrt{\tilde{K}/K}$ and
$E_{e}\rightarrow E_{e}(\tilde{K}/K)^{3/2}$, respectively.
Here, the rescaling generally depends on the rigidity of cosmic rays.

Note also that the energy densities of radiation and magnetic fields differ
by a factor of $\sim 2$--$3$ within the diffusion region $\sim 1$ kpc
\cite{Strong:1998fr}.
These fluctuations do not lead to a large dispersion of the cutoff width
as shown in Eq.~(\ref{eq:dcutdb}),
whereas the mean energy density (i.e., the cutoff energy itself)
may have an uncertainty by a factor of $\sim 2$--$3$.
Above $\sim 500$ GeV, 
the Klein-Nishina suppression of the inverse Compton cooling 
is also important, \cite{Stawarz:2009ig,Schlickeiser:2009qq}
since the optical and infrared photons dominate the radiation energy density,
while the synchrotron cooling still operates.
This effect would raise the cutoff energy by a factor of $\sim 2$,
but not affect the cutoff width considerably as long as the
energy loss time is shorter for higher energy.
The form of the source cutoff around $\varepsilon_e \siml \varepsilon_{\max}$
also affects the cutoff width of 
\beqa
\left(\frac{\Delta \varepsilon_{\rm cut}}{\varepsilon_{\rm cut}}\right)_{\varepsilon_{\max}}
\sim \frac{1}{bt \varepsilon_{\max}}
\sim 20\% \left(\frac{\varepsilon_{\rm cut}}{600\ {\rm GeV}}\right)
\left(\frac{\varepsilon_{\max}}{3\ {\rm TeV}}\right)^{-1}.
\eeqa
In addition, as we have shown in Figs.~\ref{fig:e+frac} and \ref{fig:flux},
a soft source spectrum $\alpha>2$ leads to
a round spectral shape below the cutoff,
making it difficult to measure the cutoff width.
Therefore, only under special conditions 
($\alpha \siml 2$, $\varepsilon_{\max} \simg 
5 \times \varepsilon_{\rm cut}$, 
$t_{\rm age} \simg 5 \times \Delta t$), 
a GRB-like source has a distinguishable feature of a sharp cutoff,
while, once detected, it could pindown the GRB-like nature of the source.

Our GRB-like astrophysical scenario is compatible with the diffuse 
gamma-ray background observations
because the electron plus positron flux from a GRB/pulsar is 
comparable at most to the local electron spectra (see Fig.~\ref{fig:flux})
and the locally observed electron spectra predict a minor role of the 
inverse Compton emission far below the pion decay component
as in the ``conventional model''
described in Ref. \citen{Strong:2004de}.
The constraints are less severe in the GRB case of small chance probability.
The fluctuation of the diffuse gamma-ray background due to the nonuniform 
distribution of GRBs/pulsars could be an interesting future probe.
Note that we can roughly estimate the diffuse gamma-ray flux 
from the electron flux as follows.
Gamma-rays with $\varepsilon_\gamma \sim 10$--$1000$ GeV
are produced by electrons with $\varepsilon_e \sim 100$--$1000$ GeV
via inverse Compton scattering of $\sim 1$ eV photons,
and the total gamma-ray energy created during the cooling time
$\sim 1/b \varepsilon_e$
is comparable to the electron total energy.
Since it takes only a time
$\sim 2[K(\varepsilon_e)/b \varepsilon_e]^{1/2}/c$
for gamma-rays to cross the diffusion length of electrons
while electrons stay there for the cooling time $\sim 1/b \varepsilon_e$,
the flux ratio is about the time ratio
\beqa
\frac{\varepsilon_\gamma^2 \Phi_\gamma(\varepsilon_\gamma)}
{\varepsilon_e^2 \Phi_e(\varepsilon_e)}
\sim 2 \left[\frac{K(\varepsilon_e) b \varepsilon_e}{c^2}\right]^{1/2}
\sim 0.05 \left(\frac{\varepsilon_e}{10^3\ {\rm GeV}}\right)^{2/3},
\eeqa
which is below the pion emission in the GeV-TeV region.

Similar GRBs in our Galaxy may have been observed as
mysterious TeV gamma-ray sources, the so-called 
TeV unidentified sources, which have no clear counterpart 
at other wavelengths \cite{Ioka:2004kv,Ioka:2009dh}
and/or the 511 keV electron-positron annihilation line
from the Galactic bulge \cite{Bertone:2004ek,Parizot:2004ph}.
There is evidence that GRBs predominantly occur in 
galaxies with less metals than our own \cite{Fruchter:2006py,Stanek:2006gc}.
However, no metallicity-energy correlation is confirmed \cite{Savaglio:2008fj}.
There may also be a bias that a metal-rich region tends to
have no optical afterglow due to dust absorption
(i.e., dark GRB, which is about half of all the GRBs)
and lacks host identifications and metal measurements.

In conclusion, 
the PAMELA, ATIC/PPB-BETS, and Fermi/HESS excesses
can be reproduced well if
a nearby GRB-like source, 
such as a pulsar, SN remnant, microquasar, or GRB, 
produces electron-positron pairs with
energy $\sim 10^{50}$ erg and a power-law spectral index 
$\alpha \sim 1.6$--$2.2$.
The spectra have a sharp cutoff that is similar to the dark matter
predictions.
The cutoff energy marks the source age, 
whereas the cutoff width has information on the source duration,
which may be resolved by future experiments like CALET.
Whether the source is single or multiple may be probed with
the anisotropy of electron and positron fluxes,
whereas whether the source is leptonic or hadronic
may be probed with the antiprotons \cite{Fujita:2009wk}
and secondary nuclei \cite{Mertsch:2009ph}.

\section*{Acknowledgements}
We thank M.~M.~Nojiri, T.~Moroi, K.~Nakayama, H.~Kodama, 
M.~Kawasaki, K.~Asano, and N.~Kawanaka for useful discussions.
We also acknowledge the helpful comments and suggestions from
an anonymous referee.
This work is supported in part
by Grants-in-Aid from the 
Ministry of Education, Culture, Sports, Science and Technology
(MEXT) of Japan, Nos. 18740147, 19047004, and 21684014.

%

\end{document}